Research Article

# Spin Entanglement – A Unifying Principle for Superconductors and Molecular Bonding

Paul O'Hara *

Istituto Universitario Sophia, Figline e Incisa Valdarno, FI, Italy; E-Mail: paul.ohara@sophiauniversity.org

* **Correspondence:** Paul O'Hara; E-Mail: paul.ohara@sophiauniversity.org

**Academic Editor:** Vardan Apinyan

## Abstract

The spin-statistics theorem is generalized to include quantum entanglement. Specifically, within the context of spin entanglement, we prove that isotropically spin-correlated (ISC) states must occur in pairs. This pairing process can be composed of parallel or anti-parallel states. Consequently, the article proposes using ISC states as a unifying principle to explain better Bose-Einstein condensates, the theory of superconductivity, and molecular and atomic orbitals, all of which involve a pairing process. The theoretical framework is established in sections 1 and 2. The other qualitative sections focus primarily on the experimental evidence to support the theory.

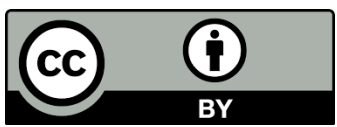

## 1. Introduction

This article proposes a unifying principle associated with rotationally invariant quantum states that brings together Bose-Einstein condensates (BEC), superconductors, and molecular bonds, all studied in Solid State Physics. We start by exploring the notion of an isotropically spin-correlated state (ISC), first introduced in [1-5], and use it to derive not only the Pauli exclusion principle but also better to understand para-statistics, Cooper pairs, and molecular orbitals. Before doing so, we note that ISC states are entangled, but not all are ISC states. More formally, if $\mathcal{H} = \mathcal{H}_1 \otimes \ldots \otimes \mathcal{H}_n$ is a Hilbert space then $|\psi\rangle \in \mathcal{H}$ is **not** entangled if $|\psi\rangle = |\psi\rangle_1 |\psi\rangle_2 \ldots |\psi\rangle_n$. This means that an unentangled state can be written as a tensor product of $n$ factors. Otherwise, $|\psi\rangle \in \mathcal{H}$ is said to be entangled. There are different graduations of entanglement based on the various degrees of factorization and the quality of the superimposed states. The interested reader is also referred to Wootters et al.'s article [6] for a more detailed discussion of the relationship between pure and mixed states in a bipartite system ($n = 2$). As noted there, "a pure state is entangled or nonlocal if and only if its state vector $\Gamma$ **cannot** be expressed as a product $\Gamma_1 \otimes \Gamma_2$ of pure states of its parts," which is in agreement with [7]. Conversely, if it can be factored as a product $\Gamma_1 \otimes \Gamma_2$ then it is not entangled, which coincides with the formal definition given above. Finally, we note that the concept of concurrence developed by Woootter et al. as a measure of pure and mixed entanglement is defined with respect to the Bell basis. As it turns out this Bell basis coincides with the ISC defined below. Indeed, for this article, we are interested only in the form of entanglement associated with ISC states.

To motivate the formal definition of an ISC state consider the two states.

$$|\psi_1\rangle = \frac{1}{\sqrt{2}}(|+-\rangle - |-+\rangle) \quad \text{and} \quad |\psi_2\rangle = \frac{1}{\sqrt{2}}(|++\rangle + |--\rangle), \tag{1}$$

where $|+\rangle = \begin{pmatrix} 1 \\ 0 \end{pmatrix}$ and $|-\rangle = \begin{pmatrix} 0 \\ 1 \end{pmatrix}$ correspond to spin-up and spin-down, respectively. Sometimes it will be convenient to drop the normalizing factor $\frac{1}{\sqrt{2}}$ in which case we write

$$|\mathbf{e}_1\rangle = |+-\rangle - |-+\rangle \quad \text{and} \quad |\mathbf{e}_2\rangle = |++\rangle + |--\rangle. \tag{2}$$

In most of the discussion that follows, the $|\psi_i\rangle$ and $|\mathbf{e}_i\rangle$ are interchangeable.

In the case of the rotation matrix

$$R(\theta) = \begin{pmatrix} \cos\theta & \sin\theta \\ -\sin\theta & \cos\theta \end{pmatrix},$$

direct multiplication gives

$$R(\theta) \otimes R(\theta)|\psi_1\rangle = |\psi_1\rangle \quad \text{and} \quad R(\theta) \otimes R(\theta)|\psi_2\rangle = |\psi_2\rangle. \tag{3}$$

In other words, the states $|\psi_1\rangle$ and $|\psi_2\rangle$ are rotationally invariant.[1]

---

[1] Note that if instead of working with $|\psi_1\rangle$, $|\psi_2\rangle$, we were to define $|\phi_1\rangle = \frac{1}{\sqrt{2}}(|+-\rangle + |-+\rangle)$ and $|\phi_2\rangle = \frac{1}{\sqrt{2}}(|++\rangle - |--\rangle)$, we would find that $R(\theta) \otimes R^T(\theta)|\phi_1\rangle = |\phi_1\rangle$ and $R(\theta) \otimes R^T(\theta)|\phi_2\rangle = |\phi_2\rangle$, where $T$ refers to

Also, in the case of each of these states, if a measurement is made on one of the components in an **arbitrary** direction then instantaneously, because of the rotational invariance and the perfect correlation between the two components, we can predict the outcome of a spin measurement on the second component in the same direction. For this reason, we say that $|\psi_1\rangle$ and $|\psi_2\rangle$ are isotropically spin-correlated states (ISC) and in that regard, note that the superimposed state $|\psi_1\rangle + |\psi_2\rangle$ is also rotationally invariant but not ISC. In addition, $|\psi_1\rangle$ is invariant not only under rotations but also under the action of any matrix $M \in SL(2, \mathcal{C})$.

As it turns out $|\psi_1\rangle$ and $|\psi_2\rangle$ are (up to a scalar) the only ISC states and therefore, we will restrict our use of the term ISC to either $|\psi_1\rangle$ or $|\psi_2\rangle$ taken separately. For this reason, they will play a vital role in the following theory, in that both of these states lay the foundation for the theory of superconductivity, Bose-Einstein condensates, and molecular bonding in chemistry. Indeed, in a certain way, each of these topics is a theme variation.

We now set above proving the uniqueness of these two ISC states. It should also be noted that the correlation among the states is broken once a measurement is performed on the paired ISC states.

**Definition 1.** *More formally, $n$ electrons are said to be isotropically spin-correlated (ISC), if a measurement made in an arbitrary direction on one of the particles allows us to predict with certainty the spin value of each of the other $n-1$ particles for the same direction.*

**Remark 1.** *Note that this is a special case of a perfectly correlated state (often called a GHZ state, see appendix). Our objective is to show that apart from $n = 2$, GHZ states are not in general ISC, which imposes a stronger condition than the perfert correlations that define GHZ states.*

**Theorem 1.** $|\psi_1\rangle$ *and* $|\psi_2\rangle$ *are the only ISC states up to isomorphism.*

Essentially, it is sufficient to show that ISC states exist only for $n = 2$. In other words, if we were to assume that three particles could co-exist in an ISC state, this assumption would lead to a mathematical contradiction. It also follows that the impossibility of three ISC particles excludes the possibility of $n \geq 3$ ISC particles.

**Proof.** Assume that there is an ISC state $|\psi\rangle$ for $n = 3$, of the form

$$|\psi\rangle = \frac{1}{\sqrt{2}}[\,|+\rangle|+\rangle|+\rangle - |-\rangle|-\rangle|-\rangle]. \qquad (4)$$

Without loss of generality, if three ISC particles share a common state then we expect to observe either $(+, +, +)$ or $(-, -, -)$, if measurements are made in the same arbitrary direction. This means that if we associate the observed spin value $s_i$ with a direction $\mathbf{a}_i$ then $P(s_i = (+, +, +)) = P(s_i = (-, -, -)) = 1/2$. Moreover, if we assume (see Figure 1) that a measurement in the direction of the vector $\mathbf{a}_i$ yields the spin values $(+, +, +)$, it also means that if we observe $+$ on one particle, then we know from the definition of the ISC state that the same $+$ value would be observed if a measurement were made in the same direction on the other two. However, it is not necessary

---

the transpose of the matrix. In other words, there is a duality principle at work which defines a $1 \leftrightarrow 1$ relation between $|\psi_i\rangle$ and $|\phi_i\rangle$. For example, if we were to define ISC states in terms of invariance under $R(\theta) \otimes R^T(\theta)$ then by duality, we could apply Theorem 1 to $|\phi_1\rangle$, $|\phi_2\rangle$. In this article we will work primarily with $|\psi_1\rangle$ and $|\psi_2\rangle$. A comprehensive discussion of this duality would require another paper.

to actually perform a second observation. In practise, we choose to make the measurement only on one particle, since by definition of an ISC state, the single measurement yields information about the other two. In this way, we do not interfere directly with the two non-measured particles. Consequently, we are free to make three independent measurements in arbitrary directions $\mathbf{a}_1$, $\mathbf{a}_2$ and $\mathbf{a}_3$, one on each of the remaining two ISC particles. For convenience, we identify the direction $\mathbf{a}_1$ with the $x$-axis and define the $z$ axis as a direction orthogonal to $x$. We will perform further spin measurements in the $x - z$ plane. Although we know a given particle spin to be $|+\rangle$ along the $x$ axis, a subsequent spin measurement along the $z$ axis of the apparatus gives $\frac{1}{2}$ probability of measuring $|-\rangle$. In general in the case of subatomic particles like electrons or protons[2], a spin measurement in the direction $\mathbf{a}$ means that the S-G device subtends an angle $2\theta$ with respect to the $x$ axis. Moreover, given that the state $|+\rangle$ has been observed with respect to the $x$ axis, then the anticipated state in the direction $\mathbf{a}$ can be constructed from the rotation $R$ and is given by $R|+\rangle = \cos\theta|+\rangle - \sin\theta|-\rangle$, and the probability of measuring $|+\rangle$ on the second particle in the direction $\mathbf{a}$ (for the subtended angle $2\theta$) is $\cos^2\theta$ and of measuring $|-\rangle$ is $\sin^2\theta$. Therefore, before actually doing any measurements or observations, we can construct a joint probability measure of the spin with respect to any two directions $\mathbf{a}_i$ and $\mathbf{a}_j$ such that $P_{ij}(+,+) = \frac{1}{2}\cos^2\theta$ and $P_{ij}(+,-) = \frac{1}{2}\sin^2\theta$. Similarly, for the ket $|-\rangle$, $R|-\rangle = \sin\theta|+\rangle + \cos\theta|-\rangle$ and the joint probabilities are $P_{ij}(-,-) = \frac{1}{2}\cos^2\theta$ and $P_{ij}(-,+) = \frac{1}{2}\sin^2\theta$. In principle, if three ISC particles exist, a sequence of spin-correlated measurements in the directions $(\mathbf{a}_1, \mathbf{a}_2, \mathbf{a}_3)$ (for the subtended angles $2\theta_1, 2\theta_2, 2\theta_3$) can be performed on the three entangled particles. Let $(s_1(\theta_1), s_2(\theta_2), s_3(\theta_3))$ represent each particle's observed spin values in the three different directions. Recall that the above-stated spin correlation implies that if any particle is measured to be in the $s_i(\theta_i) = |+\rangle$ spin state, the probability of measuring another particle in the $s_j(\theta_j) = |-\rangle$ spin-state becomes $\frac{1}{2}\sin^2(\theta_j - \theta_i)$.

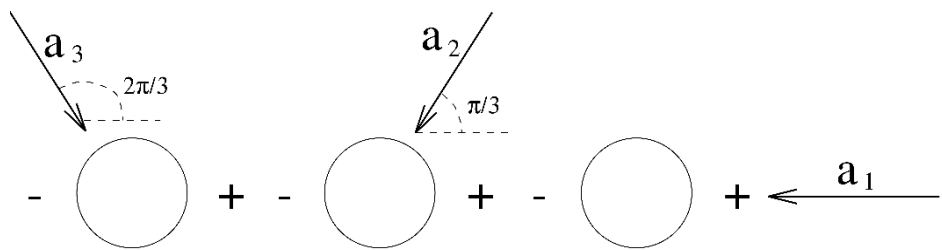

**Figure 1** Stern-Gerlach measurements in three directions on three ISC particles.

---

[2] Because of the nature of Stern-Gerlach experiments, orthogonal states are distinguished experimentally by rotations of 180° while mathematically they are distinguished by 90° rotations. Hence, a measurement in the direction $\mathbf{a}$ can be associated with the angles $(2\theta, \theta)$, where $2\theta$ correspond to the orientation of the magnetic field with respect to the $x$-axis, and $\theta$ will correspond to the value inserted into the rotation matrix in order to calculate the proper superimposed state. However, in the case of photons passing through a polarimeter, a 90° rotation is also an orthogonal state, and there is no need to associate $2\theta$ with $\theta$. In this case, we work with the same angle. The proof can be adjusted accordingly to cater for photons.

Given that $s_i(\theta_i) = |\pm\rangle$ for each $i$, there exist only two possible values for each measurement, which we associate with "spin-up" and "spin-down" respectively. Hence, for three measurements in three arbitrary directions $(\mathbf{a}_1, \mathbf{a}_2, \mathbf{a}_3)$ there are a total of 8 possible outcomes given by[3]:

$$S = \{(+,+,+), (+,+,-), (+,-,+), (-,+,+), (+,-,-), (-,+,-), (-,-,+), (-,-,-)\}. \quad (5)$$

In particular,

$$\{(+,+,-), (+,-,-)\} \subset \{(+,+,-), (+,-,-), (-,+,-), (+,-,+)\} \quad (6)$$

implies the following probability relationship:

$$P\{(+,+,-), (+,-,-)\} \leq P\{(+,+,-), (+,-,-), (-,+,-), (+,-,+)\}. \quad (7)$$

Since there are only two possible outcomes for the measurement of spin, then on summing over the second entry in the coordinates, the left-hand side of (7), can be rewritten as

$$P\{(+,+,-), (+,-,-)\} = P\{(+,.,-)\} = \frac{1}{2}\sin^2(\theta_3 - \theta_1).$$

Similarly, the right-hand side of (6) can be decomposed into the state $\{(+,+,-), (-,+,-)\} \cup \{(+,-,-), (+,-,+)\}$ from which it then follows that the right-hand side of (7) can be rewritten as

$$\begin{aligned} P\{(+,+,-), (+,-,-), (-,+,-), (+,-,+\} &= P\{(+,+,-), (-,+,-)\} + P\{(+,-,-), (+,-,+)\} \\ &= P\{(.,+,-)\} + P\{(+,-,.)\} \\ &= \frac{1}{2}\sin^2(\theta_3 - \theta_2) + \frac{1}{2}\sin^2(\theta_2 - \theta_1). \end{aligned}$$

Combining both of these results, (2) reduces to

$$\frac{1}{2}\sin^2(\theta_3 - \theta_1) \leq \frac{1}{2}\sin^2(\theta_3 - \theta_2) + \frac{1}{2}\sin^2(\theta_2 - \theta_1),$$

which is Eugene Wigner's interpretation of Bell's inequality. Taking $\theta_3 - \theta_2 = \theta_2 - \theta_1 = \frac{\pi}{6}$ and $\theta_3 - \theta_1 = \frac{\pi}{3}$ gives $\frac{1}{2} \geq \frac{3}{4}$, which is a contradiction. Therefore, three particles cannot all be in the same spin state with probability 1.

**Remark 2.** *The proof of the above theorem was worked out for $(+,+,+)$ or $(-,-,-)$ type spin correlation. To generalize the proof, suppose that the ISC particles are measured to be $(+,-,+)$ along an arbitrary measurement direction. Then, the spin outcomes in the three different directions $\theta_1$, $\theta_2$, $\theta_3$ can be written as:*

$$\{(+,-,-), (+,+,-)\} \subset \{(+,-,-), (+,+,-), (-,-,-), (+,+,+)\}.$$

---

[3] Note that the eight elements of $S$ constitute a basis for a three particle GHZ state [7], associated with perfect correlations. In this theorem, we are asking if in addition to being a GHZ state, it is also an ISC state which requires a stronger condition. As the theorem shows the answer is "no" and consequently, ISC states can only occur in pairs. Full details on the difference between GHZ and ISC states are given in the appendix to this article.

*Essentially, this means that we flipped the $|+\rangle$ to $|-\rangle$ to represent the state of particle 2. Applying the same probability argument as before but noting that $P\{(+,-,-),(-,-,-)\} = \frac{1}{2}\cos^2(\theta_3 - \theta_2)$, the inequality becomes*

$$\frac{1}{2}\sin^2(\theta_3 - \theta_1) \le \frac{1}{2}\cos^2(\theta_3 - \theta_2) + \frac{1}{2}\cos^2(\theta_2 - \theta_1).$$

*Then upon taking $\theta_3 - \theta_2 = \theta_2 - \theta_1 = \frac{\pi}{2} - \frac{\pi}{6}$ and $\theta_3 - \theta_1 = \pi - \frac{\pi}{3}$ gives as before $\frac{1}{2} \ge \frac{3}{4}$, which is a contradiction.*

**Remark 3.** *Theorem 1 and the footnote clearly indicate the uniqueness of the four Bell states in that they are also the only ISC states. It adds further weight to using the Bell states as an algebraic basis to define the entanglement measure of pure and mixed states in terms of concurrence [6].*

## 2. Electron Pairing

Historically, the distinction between Fermi-Dirac and Bose-Einstein statistics has been considered on the basis of particle spin value: a half-spin particle was thought to obey Fermi-Dirac statistics. In contrast, an integer spin particle was thought to obey Bose-Einstein statistics. However, it has been shown in [1-5] that this is not entirely correct. Pauli's original proof of the spin-statistics theorem did not apply to entangled particles. More specifically, ISC states, formed from entangled paired states, do not obey the microcausality principle, which is at the core of Pauli's proof. Indeed, the hundreds of papers (if not thousands) on the subject of 'non-locality' attest to this. The more general form of the spin-statistics theorem can be stated as follows:

**Theorem 2.** *A necessary and sufficient condition for Fermi-Dirac statistics is that the state of a system of n-indistinguishable particles be $SL(n, \mathcal{C})$ invariant.*

**Remark 4.** *For those unfamiliar with Lie group theory, it should be pointed out that $SL(n, \mathcal{C})$ is a Lie group. The set of rotations is a subgroup of this group. In light of this theorem, Fermi-Dirac statistics is best classified in terms of this group and, not in terms of the permutation group that has often been used. This Lie group and the permutation group have very different unrelated properties. The permutation group is used to define the term "indistinguishable". Different wave functions can be "indistinguishable" under different types of permutations, but the Fermi-Dirac statistics is the only eigenstate common to all elements of $SL(n, \mathcal{C})$. It is this uniqueness from the perspective of Lie groups that characterizes the Fermi-Dirac statistic.*

Various proofs have been offered in [3] and [4] and can be found in the appendix to this chapter. Since, spin value is not part of the theorem, it follows that the distinction between Fermi-Dirac and Bose-Einstein statistics is not based on spin value but rather on whether they contain indistinguishable ISC pairs or not. The ISC pairing may not be immediately obvious to someone focusing on the Fermi-Dirac state but in fact since the Fermi-Dirac statistic can be written as the wedge product of $n$ terms, it follows that any pair of them will be rotationally invariant and consequently the Fermi-Dirac statistic can be considered composed of $C_2^n$ combinations of ISC pairs. In other words, Fermi-Dirac statistics involve ISC paired states while Bose-Einstein statistics follow by breaking the paired states, meaning that their spin alignments are independent. If we consider the paired state as a singlet state then it is also possible to break the indistinguishability condition in such a way that ISC states can under the right conditions, collapse into $n$ indistinguishable or distinguishable distinct pairs of ISC states each with a total spin of 0 or 1. To distinguish both of these states, we refer to $|\psi_1\rangle (or |e_1\rangle)$ as a singlet or anti-parallel state and $|\psi_2\rangle (or |e_2\rangle)$ as a parallel state. In molecular bonding theory, we identify anti-parallel with bonding states and parallel with anti-bonding states, respectively. If we consider each paired state a new entity, then independent but indistinguishable pairs will exhibit Bose-Einstein statistics. Historically speaking, the parallel state would be called a 'vector boson' because it has a total non-zero integral spin value.

### 2.1 An Example with Two Component States

Returning to equation (1), we note that the vectors.

$$|\psi_1\rangle = \frac{1}{\sqrt{2}}(|+-\rangle - |-+\rangle) \quad \text{and} \quad |\psi_2\rangle = \frac{1}{\sqrt{2}}(|++\rangle + |--\rangle)$$

are called respectively the singlet (anti-parallel) and parallel states. It should also be noted that for any two vectors in $\mathcal{C}^2$ of the form

$$\mathbf{a} = \begin{pmatrix} a_1 \\ a_2 \end{pmatrix} \quad \text{and} \quad \mathbf{b} = \begin{pmatrix} b_1 \\ b_2 \end{pmatrix}$$

such that $a_1 b_2 - b_1 a_2 = 1$, $|\psi_1\rangle = \frac{1}{\sqrt{2}}|+-\rangle - |-+\rangle = \mathbf{a} \wedge \mathbf{b}$, which is defined by

$$\mathbf{a} \wedge \mathbf{b} \equiv \frac{1}{\sqrt{2}}(\mathbf{a} \otimes \mathbf{b} - \mathbf{b} \otimes \mathbf{a}).$$

It also follows from linearity and on combining equations (1), (2), and (3) that the state

$$c_1|\psi_1\rangle + c_2|\psi_2\rangle = \frac{1}{\sqrt{2}}(c_1|+-\rangle - c_1|-+\rangle + c_2|++\rangle + c_2|--\rangle) \qquad (8)$$

is globally a rotationally invariant state. However, it is not ISC (see definition 1).

We can combine both of these states (written in the form. $|\mathbf{e}_1\rangle$ and $|\mathbf{e}_2\rangle$) to define a Clifford algebra where the product of any two terms is rotational invariant. To see this:

**Definition 2.** *Let* $\mathbf{a}$ *and* $\mathbf{b}$ *be in* $\mathcal{R}^2$

$$\mathbf{ab} \equiv \mathbf{a} \otimes \mathbf{b} + \mathbf{a}^{\perp} \otimes \mathbf{b}^{\perp}. \qquad (9)$$

The following lemma now follows:

**Lemma 1.** *Let* $\mathbf{a}$ *and* $\mathbf{b}$ *be as above then*

$$\mathbf{ab} = \mathbf{a}.\mathbf{b}|\mathbf{e}_2\rangle + (\mathbf{a} \times \mathbf{b})|\mathbf{e}_1\rangle. \qquad (10)$$

*In other words, the product* $\mathbf{ab}$ *defines a Clifford algebra.*

**Proof:** Note that

$$\mathbf{a} = a_1 \begin{pmatrix} 1 \\ 0 \end{pmatrix} + a_2 \begin{pmatrix} 0 \\ 1 \end{pmatrix} \quad \text{and} \quad \boldsymbol{b} = b_1 \begin{pmatrix} 1 \\ 0 \end{pmatrix} + b_2 \begin{pmatrix} 0 \\ 1 \end{pmatrix}.$$

When these are substituted into the definition of the Clifford algebra and the various expressions are multiplied, the result immediately follows.

**Remark 5.** *If we define* $\langle \mathbf{a}, \mathbf{b} \rangle = \frac{1}{\sqrt{2}} (\mathbf{a} \otimes \mathbf{b} + \mathbf{b}^\perp \otimes \mathbf{a}^\perp)$ *then* $\langle \mathbf{a}, \mathbf{b} \rangle = \mathbf{a}.\mathbf{b} | \mathbf{e}_2 \rangle$ *which is a scalar product.*

In contrast, the state

$$|\psi\rangle = \frac{1}{2}(|+-\rangle + |-+\rangle + |++\rangle + |--\rangle) = \left[ \frac{1}{\sqrt{2}}(|+\rangle + |-\rangle) \right]^2 \tag{11}$$

can be interpreted as a product of two independent states from a probability perspective and can be modeled by a binomial distribution, analogous to flipping a fair coin twice. They are not independent in the sense of linear algebra. Also $|\psi\rangle$ is not rotationally invariant, although it is indistinguishable in that it is invariant under permutations.

### 2.2 Spin-1 and Two-component States

In conventional spectroscopy, a spin-1 particle can be separated into three states, which are denoted by $|1\rangle, |0\rangle, |-1\rangle$. Traditional quantum mechanics believes that each of these states should occur with an equal probability of $\frac{1}{3}$. This means that a spin-1 particle can be written as a superimposed state:

$$|\psi_d\rangle = \frac{1}{\sqrt{3}} |1\rangle + \frac{1}{\sqrt{3}} |0\rangle + \frac{1}{\sqrt{3}} |-1\rangle. \tag{12}$$

In contrast, it should be noted that the **indistinguishable** state given by equation (11) can also be re-written as the superposition of the three unit vectors. $|++\rangle, \frac{1}{\sqrt{2}}(|+-\rangle + |-+\rangle)$ and $|--\rangle$ . It takes on the form:

$$|\psi\rangle = \frac{1}{2} |++\rangle + \frac{1}{\sqrt{2}} \left( \frac{1}{\sqrt{2}} (|+-\rangle + |-+\rangle) \right) + \frac{1}{2} |--\rangle. \tag{13}$$

Written in this form, $|\psi\rangle$ is often referred to as a triplet state with each component occuring with probabilities $\frac{1}{4}, \frac{1}{2}$, and $\frac{1}{4}$ respectively. This also reflects a binomial distribution model. On comparing equations (12) and (13), we see that they are not compatible with each other. They have different probability distributions.

Which one is correct? Mathematically speaking, they represent two different and distinct models. In terms of physics, the correct model can only be decided by experiment. Indeed, the prediction that the triplet state will have a probability distribution of $\frac{1}{4}, \frac{1}{2}$, and $\frac{1}{4}$ is strongly supported by the theory of Clebsch-Gordan coefficients as noted in [1]. In other words, if a beam of composite particles, each made of two spin $\frac{1}{2}$ states that are not ISC, is passed through a Stern-Gerlach device, then it should decompose into three beams with intensity $\frac{1}{4}, \frac{1}{2}$, and $\frac{1}{4}$ and not $\frac{1}{3}, \frac{1}{3}, \frac{1}{3}$. This would be very difficult to accomplish in the case of a charged particle like a deuteron. Nevertheless, if it were experimentally verified, it would not only vindicate the approach taken here, but would also show that ISC states can help us to understand elementary particle physics better.

### 2.3 An Example with Three or More Component States

We have noted that the validity of the Pauli-exclusion principle presupposes complete indistinguishability in that a Fermi-Dirac state is anti-symmetric under permutations and completely invariant under the group $SL(n, \mathcal{C})$. Apart from the fact that the distinction between fermions and bosons is no longer based upon half-integral and integral values but upon the presence or absence of ISC coupling, it should also be pointed out that the indistinguishability conditions among particle states may be relaxed in that such conditions depend upon the initial assumptions and/or the experimental arrangement. For example, we expect that the four electrons in the beryllium atom obey $SL(4, \mathcal{C})$ Fermi-Dirac statistics while the four electrons within two separate helium atoms will obey the statistics associated with $SL(2, \mathcal{C}) \otimes SL(2, \mathcal{C})$. In other words, while the electrons within each helium atom are indistinguishable and form a pair, the different pairs respectively associated with the individual helium atoms are distinguishable from each other because the atoms are distinguishable and can also be in the same paired state. To put this in a banal way note that if two canisters of helium gas are released in separate rooms, then the atoms in distinct rooms are partially distinguishable by their separation. It follows that for beryllium, the statistics of the electron configuration contain twenty-four terms and are given by

$$|\psi\rangle_b = |\psi\rangle_1 \wedge |\psi\rangle_2 \wedge |\psi\rangle_3 \wedge |\psi\rangle_4 \,, \tag{14}$$

while in the case of the four electrons in the two helium atoms, the state contains four terms given by

$$|\psi\rangle_{2h} = |\psi\rangle_1 \wedge |\psi\rangle_2 \otimes |\psi\rangle_3 \wedge |\psi\rangle_4 \,. \tag{15}$$

Both states contain ISC-paired particles but in different quantities.

Equation (14) defines a Fermi-Dirac state for four indistinguishable particles in accordance with the Pauli exclusion principle. Indeed, it has been previously noted that a necessary and sufficient condition for the existence of a Fermi-Dirac state is invariance under $SL(n, \mathcal{C})$. What happens in the case of parallel spin states? Reviewing equation (10), which defines a Clifford algebra, we note that it is composed of an inner and an outer product. Geometrically speaking the outer product of $n$ vectors raise the grade of the multivector and corresponds to volume. It remains invariant under the action of the $SL(n, \mathcal{C})$ constitutive of a Fermi-Dirac statistic while inner products lower the dimension of the multivectors and, indeed, in the case of two vectors define a scalar, which makes it impossible to generalize it to three terms. In regular vector calculus expressions like $\langle \mathbf{a}, \mathbf{b} \rangle$, $(\mathbf{a} \wedge \mathbf{b}) \cdot \mathbf{c}$, and $\mathbf{a} \wedge \mathbf{b} \wedge \mathbf{c}$ exist and are well-defined but the inner products of three vectors are not defined. Moreover, if one were to proceed inductively and replace, for example, the antisymmetric terms in $\mathbf{a} \wedge \mathbf{b} \wedge \mathbf{c}$ with symmetric terms corresponding to parallel states, the new expression would no longer be rotationally invariant, as is the case with the standard inner product. Consequently, if they exist in nature, the behavior of parallel states will have very different characteristics to singlet states. It is also worth noting that it is quite possible, if the conditions are correct, to have $2n$ particles forming $n$ ISC paired states all in the same energy state. Moreover, as the temperature tends to 0, we can expect that these will be in the lowest energy state and can form a BEC if the crystal structure is correct and the underlying Casimir forces [8] do not break the bonding. Indeed, in order for the bonds to be broken, an energy greater than $hv$ will be needed. One of the objectives of this article is to point out that

- There are at least three categories of Bose-Einstein condensates, distinguished by the words uncorrelated states, parallel ISC states, and antiparallel ISC states. Further categories can be created by mixing these three possibilities in different proportions.
- The theory of super conductors presupposes the three categories listed above. It also follows that there are two types of Cooper pairs composed of anti-parallel and parallel states.
- In the molecular bonding theory, the orbitals associated with bonding and anti-bonding states can be identified with singlet and parallel states.

## 3. A Higher Order Viewpoint

The approach taken in this article is that the ISC states represented by $|\psi_1\rangle$ and $|\psi_2\rangle$ serve as a point of departure and can be considered a unifying factor in understanding superconductors, BEC theory, and molecular bonding in chemistry. We are neither attempting to replace various theories such as BCS nor the thermodynamics associated with calculating. $T_C$ for different types of superconductors nor are we attempting to analyze the percentage of molecules found in the BEC state inside superconductors. The objective here is to show that the different types of group structures that can be associated with ISC pairs can also serve as a unifying factor in understanding the great variety of BEC and superconducting states. The pairing alone does not explain everything and operates in accordance with the other laws of physics. Indeed, it is already apparent from the literature that the effects of temperature and pressure, electromagnetic forces, crystalline structure, geometry, and background noise associated with Casimir and van der Waal forces all play their own role in explaining and modulating the above phenomena. For example, recent experiments described in [9] and [10] show how Cooper pairs can be fine-tuned by means of Yu-Shiba-Rusinov (YSR) states. In effect, the YSR states result from modifying and interfering with the regular superconductor surroundings by using magnetic impurities. Nevertheless, suppose we also include the uniqueness of the ISC states (there are only two of them) then in that case, their addition contributes to a further refinement of our understanding, analogous to how the periodic table helped unify and explain many chemical properties that had been previously discovered and at the same time anticipated further developments.

In the context of the ISC pairs, the first thing to note is that the pairing phenomenon applies to charged particles and quantum states associated with uncharged particles such as photons and neutrons. Indeed, the whole experimental framework that was used to test Bell's inequality used singlet-state photons [11]. Neutrons in the singlet state have also been detected in large nuclei where $Z \neq N$ [12]. In other words, ISC states are a phenomenon in themselves. They obey the laws of physics but are also constitutive of a new emergent physical law that is not necessarily dependent upon force fields but rather on probability laws associated with isotropy and a higher-order symmetry related to group representation theory.

It follows that to consider spin strictly as an electromagnetic phenomenon is misleading. Charged particles with spin can acquire angular momentum through electromagnetic interactions, but the rotational invariance is not a consequence of electromagnetism. For example, singlet state photons are not susceptible to electromagnetic interactions but are isotropically polarized. Similarly, the isotropic characteristics of charged singlet states means that there is no preferred direction when measuring a charged particle's angular momentum. Often, the interaction with the measuring instrument imparts spin angular momentum.

The second thing to note is that an ISC state is primarily a geometric and group theoretical property associated with parallel or anti-parallel alignment, which emerges spontaneously in nature under the right conditions. The right conditions are multiple and depend upon the presence and/or absence of the correct temperature and pressure, electromagnetic and nuclear forces, background noise and radiation and in the case of chemistry, the rules of valency governing atomic and molecular bonding. One expalnation for the formation of ISC pairs in low-temperature superconductors, for example, is that forces that might otherwise disrupt and tear apart the ISC alignment are diminished so that paired states can form naturally. Something similar occurs in high-energy superconductors where in the case of (almost) perfect crystals, the background environment is uniform throughout and constructed so that the background interference that might prevent the formation of the pairs is mitigated. Other things being equal, the right background provides for the spontaneous development of the pairs.

We might ask what causes the spontaneous alignment. It cannot be a force field otherwise, we would have to have an exchange particle mediating the interaction beyond the light cone. Certainly, phonons can mediate the forming of pairs when particles are close to each other, but once the pairs are formed and remain intact, the phonons have at best, no role to play and at worst may be a factor in breaking the entanglement. Two non-interacting particles can be aligned because they are rotationally invariant and non-local and violate microcausality. Again we might ask why rotational invariance occurs? The initial alignment may have been mediated by close-range interactions between the particles, such as phonons or the Casimir effect. Nevertheless, even if the particles come close enough to interact, that does not explain the rotational invariance. External (and also internal) forces bring the two particles together like interlocking pieces of a jigsaw puzzle, allowing the ISC state to be formed, but that does not explain why such states exist in the first place. ISC states remain a mystery associated with Godel's theorem in that no mathematical system equals its own completeness. Higher-order viewpoints require additional axioms and/or hypotheses and/or the suspension of old axioms. ISC states occur because, mathematically, they are permitted within the Hilbert space theory of quantum mechanics. It is part of the mystery of existence akin to asking why spacetime is best modeled by a four-dimensional pseudo-Riemannian manifold. It also reflects Leibnitz's question, "Why is there something rather than nothing?"

### 3.1 Condensate States

The spin-statistics theorem presented here does not depend on the spin's value. Fermi-Dirac statistics is a consequence of indistinguishable spin pairing, not value. As the indistinguishability is relaxed various forms of para-statistics emerge. For example, we have already noted that in the case of four electrons within beryllium (Be), we can expect $SL(4, \mathcal{C})$ statistics, while if the four electrons are associated with two independent helium (He) atoms, then the four electrons will obey $SL(2, \mathcal{C}) \otimes SL(2, \mathcal{C})$ statistics, while the three electrons of Lithium and the isolated electron of hydrogen obey $SL(3, \mathcal{C}) \otimes SO(2, \mathcal{R})$. Since spin value has no role to play, we might postulate that Bose-Einstein condensation is possible for electrons. If we consider a superconductor as a moving condensate (see below), we might consider a condensate as a superconductor from the perspective of the rest frame.

In terms of parallel pairing, the existence of distinguishable pairwise $SO(2, \mathcal{R})$ symmetry states mean that the isotropy is defined with respect to two 2-dimensional planes and presuppose a weaker energy bond. Hence, parallel pairwise bonds are easier to break and, consequently, will be more likely to undergo interactions with its environment, including chemical ones. There are different ways that parallel pairing can take place depending on the inclination of the respective planes. In practice, the actual pairing will depend upon the physical conditions under which they are created. For example, if two charged particles are parallel entangled, it would be reasonable to expect the ISC states lie in planes perpendicular to the direction of electrostatic attraction or repulsion. In particular, if they form a condensate, they will exhibit parallel planar alignment in every direction at once by virtue of the isotropy associated with the entanglement. Moreover, suppose Casimir forces are introduced along the line joining the center of mass of the two objects. In that case, the interplay between Casimir forces will always add an attractive component that enhances or diminishes electromagnetic forces.

### 3.2 Three Possible Condensate States

The essence of BEC is that particles, atoms and/or molecules are all in the same lowest energy state. The BEC may appear as a macroscopic object, although it comprises a conglomerate of microscopic entities. Moreover, the closer these entities resemble macro objects, the more they will appear to be (almost) at complete rest with respect to the laboratory. However, appearances can be misleading, and with this in mind, we distinguish five possibilities:

1. An aggregate of independent atom molecules is brought to rest (almost). No pairing is involved. The first ever BEC state was formed from loosely bound rubidium ($Rb^{87}$) atoms. Although they are at rest, they are still referred to as a gas because of their low density. $Rb^{87}$ atoms are repulsive, and consequently the distance separating them stems from an equilibrium between the repulsive forces and the Casimir or Van der Waals attractive forces, first discovered by London (1930), which are usually attributed to zero-point energy (p.99) [13]. Moreover, $Rb^{87}$ is technically not a boson in the "older" sense of the term in that the condensate is a collection of distinguishable atoms. In contrast, a proper Bose-Einstein statistic would have required that they be indistinguishable. Consequently, such a state should be impossible. Distinct atoms are distinguishable by definition. Indeed, Einstein predicted that BEC could not be created for that reason. With the new formulation of the spin-statistics theorem, there is nothing to prevent paired or unpaired states from being brought to rest, at least in principle. Whether or not it happens depends on technology and the relative strength of the vacuum forces. When Cornell and Wieman created the first BEC, they cooled the atoms by slowing them down using laser beams and further reducing the temperature by removing the remaining moving atoms with powerful magnets. What remained were about 2000 atoms in a non-dynamic equilibrium that lasted for 15-20 seconds in a diameter of 20 microns ($10^{-6}$m), which corresponds to a particle separation distance of $\sim 10^{-8}m$. In a similar experiment posted in [14], the number of $Rb^{87}$ atoms forming the BEC state are between $10^5 - 10^6$, lying within a radius $0.1mm - 0.25mm$. It should also be pointed out that Einstein did not predict the existence of BEC states but the opposite in that he "concluded that the theory provided a paradox because it predicted a state with indistinguishable particles occupying this same volume and as Einstein said 'but this appears to be as good as impossible.' "[15] Finally, if someone were observing this condensate from the sun, they could claim that they are observing a superconductor of distinct atoms moving in a gravitational field.[4] Mathematically speaking, the total energy of the system in the rest frame is given by

$$E = T + V = V = \frac{1}{2} n \hbar \omega, \tag{16}$$

where $n$ corresponds to the number of atoms in the condensate and $V(r)$ is the potential energy due to the Casimir force of the vacuum and the lattice structure.

2. The second type of condensate occurs when elementary particles or atoms form pairs during the cooling process, although these pairs may already exist as for example, in the case of molecular bonds at room temperatures or "Pre-formed Cooper pairs in copper oxides and $LaAlO_3 - SrTiO_3$ heterostructures." [16] In other cases, the pairing will occur below a specific critical temperature $T_c$. This is called BCS-BEC crossover. One may think of the process of cooling room temperature. $O_2$ to a BEC state. This, too, involves pairings associated with molecular bonding. Indeed, while writing this article, in the case of $Rb^{87}$ we predicted that if the density between the atoms is increased, that should form $Rb_2^{87}$ molecules, only to be pleasantly surprised during a literature search that such molecules have already been discovered [17]. However, it is essential to note that the diatomic molecules are not forming ISC states. It is the electron bonds holding the atoms together that form parallel or antiparallel ISC states.

3. This last example begs the question as to whether we can have BEC states of elementary particles. This brings us back to Einstein's original observation that "the theory provided a paradox because it predicted a state with indistinguishable particles occupying this same volume and as Einstein said 'but this appears to be as good as impossible.' "[15] Nevertheless, given that BEC condensates of hydrogen atoms have been created, it is not difficult theoretically to imagine the existence of a BEC state of hydrogen ions, where the electrostatic repulsion of the protons is counterbalanced by the van der Waals' and Casimir attractive forces. Of course, experimentally speaking, this would be a lot more difficult to achieve given the repulsion between the protons and that the critical temperature. $T_c$ is inversely proportional to mass [18] (see equations 2.4.4 and 2.4.11), but the interaction should also allow for indistinguishable spin states to form a Femi-Dirac Condensate (FDC) with a frequent making and breaking of bonds with the nearest neighbors forming a dynamic equilibrium in the Fermi sea. This would be best performed in some type of trap. Indeed, on reading Bardeen's description of Cooper Pairs, one has the impression that their behaviour is more akin to Fermi-Dirac statistics. Bardeen notes that "the idea of paired electrons, though not fully accurate, captures the sense of it" [19]. More precisely, about Bardeen's quote, Delin and Orlando note that "we should not think of Cooper pairs as tightly bound electron molecules. Instead, there are many other electrons between those of specific Cooper pairs, allowing the paired electrons to change partners on a time scale of $h/2\Delta$, where $h$ is Planck's constant" [20] and $2\Delta$ represents the binding energy of the pairs. This making and breaking of the Cooper pair bonds is more akin to the Fermi-Dirac statistic based on the more general form of the spin-statistics theorem (see Appendix). Bardeen would have been unaware of this new form of the theorem and consequently, associated the pairing phenomenon with Bose-Einstein statistics. Nevertheless, his own quotation about the nature of the pairing would better fit Fermi-Dirac statistics as described above. One would hope that this could also be extended to electron clouds. Moreover, given the small mass, we would expect the Casimir or Van der Waals attractive forces to facilitate the making and breaking of the electron pairs. It will be a true FDC.

---

[4] In practise, the BEC as prepared in the laboratory actually is free falling in a gravitational field.

4. We have postulated the existence of trapping hydrogen ions to form a BEC state. It should also be noted that the evidence suggests that the BEC between multiple electrons is created within superconductors. Quantum mechanically, this can be described by a series of quantum standing waves (see below). No magnetic field will be present because there is no relative motion between them. Moreover, if they were to move as a block, as seems to be the case with superconductors, then there would be no inner magnetic field, but the block would induce a magnetic field outside on and around the surface.

5. Finally, we note there is no reason in principle that BEC state cannot involve parallel pairing. Indeed, the concept of anti-molecular bonding in chemistry may be seen as a potential BEC state. Essentially, the state can be created by forming the parallel ISC states and then cooling to a BEC state. An example of this might be found when considering $U_7Te_{12}$ where parallel states seem to form very close to absolute zero [21]. Since these states exhibit $SO(2,R)$, Invariance will be easier to create if the BEC pairings are all in the same plane or, better still, perpendicular to a fixed plane.

### 3.3 Superconductors

We define a superconductor as a moving BEC. For example, we referred to the $Rb^{87}$ BEC as a superconductor moving under gravity as seen from the perspective of the sun. It moves as a block, with no relative motion between the particles and without electromagnetic resistance. From the standpoint of general relativity, we could say that it is moving on a geodesic and consequently free-falling in a quantum (and not classical) vacuum, mediated by the presence of Casimir forces, which either enhance or impede the motion. It is difficult to know how these forces work and what their causes are. Perhaps gravity itself is a manifestation of these forces.

This being the case, if a BEC state is defined in Minkowski space, then its collective motion will be Lorentz invariant. In addition, an internal higher-order symmetry is associated with the ISC pairs. Specifically in the case of a BEC composed of singlet states, they form a Lorentz invariant higher-order symmetry [2], while in the case of parallel pairing, there is an $SO(2,R)$ invariant symmetry. This latter symmetry is more limited because the ISC correlated planes are perpendicular to the direction of motion or to express it in algebraic language, $SO(2,R)$ is a subgroup of the Lorent group. Given these characteristics of superconductors and the fact that bounded ISC particles move freely without resistance inside a conductor, we might ask what characteristics of the vacuum field within a superconductor explain such a motion.

### 3.4 The Field Inside a Superconductor

The two pertinent characteristics to help understand the field inside the ideal superconductor is that particles are considered BEC condensates formed of paired ISC states that move collectively against a backdrop of a perfect crystal, corresponding to periodic motion. This suggests that there is also an isotropic EM-field within the superconductor, which is analogous to the field in free space. The motion is best modeled by an (infinite) series of 3-dimensional harmonic oscillators [22]. This being so, following Milonni [13] we note that the field Hamiltonian is of the form

$$H_f = \sum_{k_\lambda} \hbar\omega_k \left( a_{k_\lambda}^\dagger a_{k_\lambda} + \frac{1}{2} \right) \tag{17}$$

such that

$$\left[ a_{k_\lambda}(t), a_{k_\lambda(t)}^\dagger \right] = \delta_{k,k'}^3 \delta_{\lambda\lambda'},$$

and $\lambda$ refers to the bi-polarity of the spin field. Moreover, the vector potential in free space is given by

$$A(\mathbf{r}, t) = \sum_{k_\lambda} \left( \frac{2\pi\hbar c^2}{\omega_k V} \right)^{\frac{1}{2}} [a_{k_\lambda}(t)\exp(i\mathbf{k}.\mathbf{r}) + a_{k_\lambda}^\dagger(t)\exp(-i\mathbf{k}.\mathbf{r})]\mathbf{e}_{k_\lambda} \tag{18}$$

from which it follows that

$$\mathbf{E}(\mathbf{r}, t) = \sum_{k_\lambda} \left( \frac{2\pi\hbar\omega_k}{V} \right)^{\frac{1}{2}} [a_{k_\lambda}(t)\exp(i\mathbf{k}.\mathbf{r}) - a_{k_\lambda}^\dagger(t)\exp(-i\mathbf{k}.\mathbf{r})]\mathbf{e}_{k_\lambda} \tag{19}$$

and

$$\mathbf{B}(\mathbf{r}, t) = i \sum_{k_\lambda} \left( \frac{2\pi\hbar c^2}{\omega_k V} \right)^{\frac{1}{2}} [a_{k_\lambda}(t)\exp(i\mathbf{k}.\mathbf{r}) - a_{k_\lambda}^\dagger(t)\exp(-i\mathbf{k}.\mathbf{r})]\mathbf{k} \times \mathbf{e}_{k_\lambda}. \tag{20}$$

Note that, $c^2 = (\epsilon\mu)^{-1}$ corresponds to the velocity of light inside a conductor with permittivity $\epsilon$ and permeability $\mu$, which is different to the velocity of light in vacuum. Using the definition $P = \frac{1}{4\pi c}\int_V d^3r(E \times B)$, we obtain

$$P = \sum \hbar\mathbf{k} \left( a_{k\lambda}^\dagger a_{k\lambda} + \frac{1}{2} \right). \tag{21}$$

It now follows that $[P, H_f] = 0$ where $H_f$ is the field Hamiltonian, which means that the linear momentum is a constant in the direction of motion. In what follows below, we assume that the direction of motion is the x-axis.

With this in mind, we can now apply the above to anti-parallel and parallel ISC states. Note that the bipolarity term represented by $\lambda$ will be absorbed into the ISC state. If these states are oscillating in unison and also in movement because of an (electric) potential difference, this will lead to extending the 2-dimensional Hilbert space $\mathcal{H}$ to the space $\mathcal{H} \otimes L^2(r, t)$. This enlarged state will experience a twofold Lorentz invariance. The first is between the pairs and the second with respect to the the individual motion of each electron in the superconductor relative to the direction of the potential and subjected to Maxwell's equation. As previously explained, the ISC states may not actually be spinning, and if they are, they must be in such a way that the entanglement is preserved, and in the case of the singlet state, the sum of their angular momentum is 0. Moreover, suppose there is equal spacing between the pairs as if they are a (virtual) lattice. In that case, the pairs within the potential will not experience any relative motion with respect to each other, in which case there will be no induced magnetic properties. On the other hand, if this moving lattice is itself in relative uniform motion with respect to the lattice points of the conductor, again, there will be no induced electromagnetic currents since $div B = curl B = 0$. Indeed, in the rest frame of the electron lattice $B = 0$.

Equivalently, if we let $\xi$ (a constant) be the separation distance between the center of mass of adjacent ISC pairs of electrons and $v$ the constant velocity (along the $x$ axis) with respect to the fixed points of the lattice, then as quantum particles the wave function with respect to the n-nodes of the lattice can be expressed as $\psi = \sum_n \psi_n$ such that:

$$\psi_n(x,t) = \frac{1}{s_o\sqrt{2}}\left[\exp\big(ik(x-vt)\big)\exp(-2i\hbar^{-1}m^*cs)(|+-\rangle - |-+\rangle)\right], \text{where } x-vt=n\xi, k=\frac{2\pi j}{\xi}, j,n \text{ are integers} \quad (22)$$

for the singlet state, while for the ISC parallel state,

$$\psi_n(x,t) = \frac{1}{s_o\sqrt{2}}\left[\exp\big(ik(x-vt)\big)\exp(-2i\hbar^{-1}m^*cs)(|++\rangle + |--\rangle)\right], \text{where } x-vt=n\xi, k=\frac{2\pi j}{\xi}, j,n \text{ are integers.} \quad (23)$$

The above wave functions have been derived directly from the kinematics of superconducting motion, assuming that Cooper pairs are ISC correlated, that the separation between the two electrons of the ISC (Cooper) pair is constant (dependent upon $k$) and that separation between different ISC pairs ($\xi$) is also constant. Moreover, as quantum particles we expect for the singlet (anti-parallel) and parallel states, respectively that in addition to the uniform motion, there will be a zitterbewegung vibration such that $m* \approx 2m$, with $m$ the electron mass and $m*$ the mass of the ISC pair. We note that the zitterbewegung term $\exp(-2i\hbar^{-1}m^*cs)$ can be interpreted as a periodic isotropic vibration corresponding to a pair of synchronised particles such that $s \in [-s_o, s_o]$ with $s_o$ being the maximum wavelength of the vibration of each particle.

### 3.5 The Effects of Magnetic Fields on Superconductors

It follows that the magnetic dipole characteristics of each entangled state, whether in the parallel or anti-parallel state, depend not on the entanglement per se but upon the characteristics of the external magnetic field in which it is immersed. As long as the set of entangled states is in a superconducting state, there is no relative motion between them, and consequently, there is no interior magnetic field (meaning a constant magnetic potential). This will be manifested as the expulsion of a magnetic field (The Meissner Effect). In other words, if the magnetic field is not strong enough to break ISC bonds and/or the lattice structure, it will be eliminated from the superconductor. Magnetic properties are manifested in charged particles that are accelerating. For quantum phenomenon, this means that there is a minimum energy $E = \hbar\omega_{min}$ required to accelerate or remove a particle from a lattice. Below this threshold, magnetic fields will not appear in that they can be considered as having a constant magnetic potential $\mathbf{A}$ such that $curl \mathbf{A} = 0$.

Indeed, often, the presence of the magnetic field will destroy the entanglement by breaking the symmetry between the entangled particles and replacing it with a polarization effect due to the induced magnetism, although in some cases, it might actually enhance the entanglement (as in the case of some molecular bonding in chemistry). In fact, it appears both situations are possible theoretically (and experimentally) speaking. For example, an external magnetic field will induce magnetic dipoles properties into each pair of particles, causing them to further align or become independent. If the magnetic dipoles associated with each particle are aligned on a line, one behind the other like compass needles, then their N-S poles will attract each other but in two different ways. If the magnetic field is strong enough, it can break the ISC coupling and create a string of N-S dipoles that align, having the overall characteristics of one large ferromagnet.. However, this characteristic will usually disappear when the external magnet is switched off. On the other hand, if the field does not break the ISC coupling but causes them to rotate in unison, then pairs will behave like diamagnets (especially in parallel cases) and exhibit properties associated with rotating superconductors.

Their behavior will be affected by the strength of the magnetic field in which they are placed and their orientation. If a superconductor current is moving along the positive $x$-axis and the magnetic field is placed at an angle $\theta$ then the magnetic characteristics will depend upon both the strength and the alignment angle employed [21, 23]. If we have a strict polarization effect and $n$ dipoles are lined up on a line then there will be a single N-S dipole effect due to cancelations. If the magnetic field rotates, cancellation effects may no longer exist, and the electrostatic repulsion between ISC parallel states may increase and even flip polarity. Moreover, if the magnetic field is strong enough, it will pull them apart. For example, if all the magnetic dipoles are in equilibrium and parallel to each other (as well as parallel to the field) then as the field strength increases the magnetic repulsion between the particles will increase until they are eventually independent of one another. On the other hand, in the case of a singlet state, if the induced magnetic dipoles are placed side by side analogous to magnetic dipole moments created by two parallel wires carrying a current in the same direction, then they will not only attract each other but will also cancel each other out. In reality, as experiments with superconductors have shown, the possibilities are almost endless. Temperature, Casimir forces, electromagnetic properties, and the geometry of the potential superconductor all play a role, and to tame these possibilities means operating with simple models. Also, the critical role of temperature $T_c$ in aiding or impeding superconductivity has been noted but it alone cannot explain everything in that $T_c$ depends upon the chemical composition of the material, its geometry (think of the different allotropes of carbon) and whether the pairing is parallel or anti-parallel.

Finally, we note that in principle, one does not need pairing to have superconductivity. However, given the natural tendency of particles to pair (which is the thesis of this manuscript), in order to create non-pairing super conductors, one will need to have large spatial separation between particles to prevent the pairs from forming. For example, as already mentioned, when $Rb^{87}$ was originally used to form a BEC condensate the spatial separation of the atoms was on the macroscopic level but when they were brought closer together then transitioned into $Rb_2^{87}$ molecules, which means that the unbounded electrons came together to form ISC orbitals. Also, as previously noted, the BEC is composed of $Rb^{87}$ relative to the sun, could be considered as superconducting under gravity. The challenge would be to do the same thing with electromagnetic forces.

### 3.6 Molecular Bonding

The basic thesis is that there is a natural tendency for elementary particles like electrons to form ISC pairs and indeed these bonds can be long-ranged once they are formed. Geometrically speaking, the pairing is analogous to interlocking pieces of a jigsaw in that the particles have to be close enough together to be matched. In other words, we claim that although the pairs can remain entangled over a long range, they are formed at close range through electromagnetic and/or strong interactions, and Casimir forces. Once formed, the way in which these pairs are distributed in spacetime depends on how long it takes for decoherence, which in turn depends upon temperature, electromagnetic and other forces. It is precisely these variable factors operating randomly at times that give rise to quantum statistics.

Chemical elements are a natural domain for forming ISC entangled pairs. The simplest way to form such pairs is within atomic and/or molecular orbitals (meaning having the same energy and orbital angular momentum) but it is not strictly necessary. As modern chemistry has demonstrated, one can also have paired particles involving different orbitals, such as ortho- and para-oxygen [24]. Another example can be found by considering a positronium state which is formed by a positron and an electron. In this positronium state, the two leptons share the same atomic-like orbital, and their spins may be correlated as parallel or anti-parallel, corresponding to para-positronium and ortho-positronium variants. Moreover, this positronium example demonstrates that spin correlation between two leptons is generally not restricted to only the anti-parallel case, even when they share the same orbital. By analogy, we consider delocalized electron pairing with parallel correlated spins. This analogy is appropriate for delocalized electrons because their wave function is not centered around any nucleus.In this case, two delocalized electrons form a pair via direct lepton-lepton interaction mediated by Casimir forces, similar to positronium.

In either case, we might ask what causes the atomic and molecular orbitals. The Dirac equation predicts the existence of spin states, but it does not predict ISC pairs. That said, we maintain that the quantum vacuum that permeates all of spacetime and exists between the nucleus and the electrons in the atom is mediated by the Casimir forces, which enable the orbitals to form. Moreover, there seems to be a preference for singlet state pairing in that if momentum is induced into the pairs the singlet state will have spin angular momentum 0, in contrast to weak molecular bonds where parallel pairing takes place but has a spin angular momentum of $\pm 1$. In molecular chemistry, we associate these two different ISC states with anti-parallel and parallel bonding.

In general, atoms are a natural domain for the formation of ISC pairs, and oftentimes, the indistinguishability conditions within the atom and molecule manifest themselves as Fermi-Dirac statistics. Nevertheless, as the above theory has aptly demonstrated, Fermi-Dirac statistics is not an absolute but rather highlights the pairing process when complete indistinguishability is involved and also highlights the fact that indistinguishability conditions can be relaxed. For example, the molecular bonds associated with a molecule of ammonia ($NH_3$) are formed by three hybrid orbitals in which all the electrons are in the same state.

## 4. Conclusion

As stated in the beginning, the objective of this article is to show that ISC pairing is a unifying principle when it comes to our understanding of BEC states, superconductivity and molecular bonding theory. Moreover, an ISC pair should be considered a single entity. It reflects an Aristotelian principle that the whole is more than the sum of its parts. Also, the Pauli exclusion can be derived by assuming that the particles forming ISC states are indistinguishable. Likewise, the conditions of indistinguishability can be relaxed, and many types of para-statistics can emerge. Also, the Bose-Einstein statistics presuppose complete indistinguishability with the quantum state but without ISC pairing.

The other key factor in understanding the pairing phenomenon is that it functions as an independent hypothesis within the system. It is caused by the geometric structure associated with spin. It is not derivable because once the Hilbert space structure on which quantum mechanics is constructed is permissible, the ISC state becomes possible. It is not caused by any external forces. It is an independent state constitutive of the geometric properties of spacetime. However, it is subjected to the laws of physics and will respond to other physical phenomena such as temperature, electromagnetic, and nuclear forces. The initial creation of these states appears to be mediated by Casimir forces that allow the original independent states to interact and combine non-linearly to form rotationally invariant states. Nature forms them because there is no reason why they should not exist, given the structure of spacetime. They are an emergent phenomenon.

In terms of experimentation, the fact that only two ISC states are theoretically possible and that Cooper pairs and other parallel pairing phenomena are associated with superconductors cannot be a coincidence. It would seem that superconductor experiments have provided ample evidence of the existence of ISC pairs but have not fully explained them. Here, we see that the mathematical structure of quantum mechanics defined over a tensor product of two Hilbert Spaces provides the theoretical explanation and is profoundly linked to entanglement manifested as ISC states. Moreover, the theory not only explains intuitively the Pauli exclusion principle but also predicts that experimentally, if a beam of spin 1 composite particles in the triplet state is passed through a Stern-Gerlach device, it should divide into the three separate spins states +1, 0, -1 with probabilities $\frac{1}{4}, \frac{1}{2}, \frac{1}{4}$ and not $\frac{1}{3}, \frac{1}{3}, \frac{1}{3}$ as is currently believed. Admittedly, the experiment becomes very complicated if these spin 1 particles are charged (like deuterons).

Also, based on the notion of ISC states, it follows that, in principle, it should be possible to create superfluids composed of proton Cooper pairs or, in other words, of paired hydrogen ions. This is reinforced by the recent announcement of simulated Cooper pairs using Strontium atoms [25] and [26].

Finally, the role of Casimir forces has been hinted at as a fundamental backdrop that mediates many quantum interactions, and indeed, given its overall tendency to bring bodies together, one might also wonder if it is not the basic cause of gravity. It is a question that remains open and suggests an exciting way forward for future physics.

**Appendix: Entanglement, Perfect Correlations and ISC States**

The introduction to this article pointed out that ISC states are rotationally invariant and perfectly correlated. They are two different concepts, and the reader might be wondering how ISC states generally differ from other perfectly correlated states (usually referred to as GHZ states). To motivate this section, the author will refer to the famous paper "Bell's theorem without inequalities" by Greenberger, Horne, Shimony, and Zeilinger [7]. For simplicity, we will use the acronym GHSZ when referring to this article.

Following GHSZ, an entangled state is such that "it cannot be written in any way as a product of single-particle states"(p1132). In other words, when a state is entangled, it cannot be written in the form

$$|\psi\rangle = |\psi_1\rangle|\psi_2\rangle \dots |\psi_n\rangle.$$

For example, if $|\psi_1\rangle$, $|\psi_2\rangle$ and $|\psi_3\rangle$ combine to form an unentangled state, then its tensor product can be written as

$$|\psi\rangle = |\psi_1\rangle|\psi_2\rangle|\psi_3\rangle.$$

In contrast, there are different degrees of entanglement. For the three states $|\psi_1\rangle$, $|\psi_2\rangle$ and $|\psi_3\rangle$:

$$|\psi_1\rangle(|\psi_2\rangle|\psi_3\rangle + |\psi_3\rangle|\psi_2\rangle)$$

is an entangled state as is

$$|\psi_1\rangle|\psi_2\rangle|\psi_3\rangle + |\psi_2\rangle|\psi_3\rangle|\psi_1\rangle + |\psi_3\rangle|\psi_1\rangle|\psi_2\rangle.$$

In general, if we are in the space $C^3$, defined over the complex numbers, a threefold entangled state can be written as a linear combination of $3^3 = 27$ (complex) terms in the form

$$\sum \alpha_{ijk}|\psi_i\rangle|\psi_j\rangle|\psi_k\rangle,$$

where each $i, j, k, \in \{1,2,3\}$. From a mathematical point of view, depending on the values of $\alpha_{ijk}$ there are an infinite number of three-particle entangled states. In practice, we focus on certain specific states of interest in physics. For example, if $i \neq j \neq k$, then it will reduce to a state of only six terms, and if we further require that it be composed of indistinguishable terms, it will reduce to a Fermi-Dirac state. It remains an open question whether every possible mathematically entangled state can be physically realized.

If $i, j, k$ are restricted to only two values, then the threefold entangled state can be written as a linear combination of $2^3 = 8$ terms, as with the GHZ state defined by equation (G1) of GHSZ. It is given by

$$\frac{1}{4}[(1 - ie^{i(\psi_1+\psi_2+\psi_3)}|d\rangle_1)|e\rangle_2|f\rangle_3 + (i - e^{i(\psi_1+\psi_2+\psi_3)}|d\rangle_1)|e\rangle_2|f'\rangle_3 \qquad (24)$$

$$+\left(i-e^{i(\psi_1+\psi_2+\psi_3)}|d\rangle_1\right)|e'\rangle_2|f\rangle_3+\left(-1+ie^{i(\psi_1+\psi_2+\psi_3)}|d\rangle_1\right)|e'\rangle_2|f'\rangle_3 \quad (25)$$

$$+\left(i-e^{i(\psi_1+\psi_2+\psi_3)}|d'\rangle_1\right)|e\rangle_2|f\rangle_3+\left(-1+ie^{i(\psi_1+\psi_2+\psi_3)}|d'\rangle_1\right)|e\rangle_2|f'\rangle_3 \quad (26)$$

$$+\left(-1+ie^{i(\psi_1+\psi_2+\psi_3)}|d'\rangle_1\right)|e'\rangle_2|f\rangle_3+\left(-i+e^{i(\psi_1+\psi_2+\psi_3)}|d'\rangle_1\right)|e'\rangle_2|f'\rangle_3]. \quad (27)$$

These states are particularly interesting in that they represent the evolution of the GHZ state given by

$$|\psi\rangle=\frac{1}{\sqrt{2}}[|a\rangle_1|b\rangle_2|c\rangle_3+|a'\rangle_1|b'\rangle_2|c'\rangle_3\,, \quad (28)$$

where $|a\rangle_1\to\frac{1}{\sqrt{2}}[|d\rangle_1+i|d'\rangle_1]$ and $|a'\rangle_1\to\frac{1}{\sqrt{2}}[|d'\rangle_1+i|d\rangle_1]$. It should be apparent from this transformation that $|\psi\rangle$ is not rotationally invariant, even when $\phi_1+\phi_2+\phi_3=0$, whereas when projected into a two-dimensional state with $\phi_1+\phi_2=n\pi$ it is rotationally invariant under $(R,R^T)$, where $R$ is a two-dimensional rotation matrix.

It should also be noted that equation (28) is similar in form to equation (4). Indeed, if the ISC state existed, it would be a particular case for a GHZ state. However, there is a difference. The general GHZ state for three particles presupposes that if we measure (observe) the states $|a\rangle$ and $|b\rangle$ then $|c\rangle$ would be determined by the other two. However, in the case of an ISC state, we require that the measurement of **one** state would determine the other two. In other words, if we were to observe $|a\rangle$ then both $|b\rangle$ and $|c\rangle$ would be determined in the same direction of measurement. The authors of GHSZ paper seem aware of this. In their Appendix A, they point out and prove the rotational invariance of the singlet state (equation (A3)). Moreover they imply (although they do not develop it further) that there is a difference between ISC states (my terminology) and "a rotationally invariant **mixture** of product states, which will not yield correlations as strong as [the singlet] $|\psi\rangle$ does." They proceed to give an example in Appendix B. Finally, we note that in terms of GHZ states, rotational invariance is only applied to the two-particle singlet state (paired qubits). There is no reference to other GHZ states being rotationally invariant; instead they refer to polarized states. The Fermi-Dirac state defined in Theorem 3 (below) is rotationally invariant but is only an ISC state for n = 2.

It might seem like quite a tall order to expect one measurement (observation) to yield $n$ pieces of information, and certainly when it comes to quantum mechanics, it is not possible. However, if hidden variables in Bell's sense (or Einstein's sense) existed, then it would likely be possible to have $n$-ISC states for $n\geq 2$. To see this, consider by way of analogy, the state of the circumference of $n$ identical circles, all of radius $r$. If $r$ is unknown then by measuring the radius of one circle, we will know the radius of all the others. Indeed, because the radius constitutes a parameter of the circle, knowing the parameter's value allows us to independently draw as many circles as we wish of the same radius. It is sufficient to choose $n$ distinct center points and draw $n$ identical circles of radius $r$ with the compass. If hidden parameters had existed in Einstein's sense, then in principle by knowing the characteristics of the parameters for each spin state, it would be possible to have $n$ isotropically independent spin states. In other words, the correlations would be a mere epiphenomenon in that one does not need to know about correlations in order to make $n$ independent copies. It is sufficient to repeat the same rules over and over again, without in any way referencing the previous construction. In the case of the $n$ circles of radius $r$, any pair of circles (states) is a mirror images one of the other, but that information is not necessary to construct the $n$ independent circles. In this case, the correlation between them is a secondary consequence of the construction. In contrast, the proof of Theorem 1 shows that it is impossible to have three or more ISC states; in the case of two, there is no hidden parameter to define the twin-like quality of the ISC states. They are intrinsically correlated by mathematical group properties that only hold for paired states. Consequently, ISC states upto isomorphism must occur in pairs and are characterized by correlations and not hidden parameters. ISC correlations are something completely new.

It should also be apparent from the proof of Theorem 1 that we are not saying that GHZ states given by equations (28) and (4) do not exist. We are saying that they do not exist as rotationally invariant states nor as ISC states. Indeed, GHSZ refers to the singlet state as a rotational invariant state (c.f. Sections 2 and the Appendices A and B). In contrast, they refer to GHZ states composed of three (or more particles) as exhibiting "polarization correlations" (p.1138). In other words, polarization refers to correlations in specific directions, while ISC states are correlated simultaneously in all directions. The thesis of this article is that it is precisely the two ISC states that are formed when materials superconduct. Apart from the two ISC states discussed in this article, all other GHZ states exhibit some form of polarization, which can be realized in the laboratory in different ways according to the experimental apparatus used to produce such states. Section VI of GHSZ provides an experimental example of how such a state might be obtained in 'real experiments.' The ISC states are uniquely rotationally invariant and perfectly correlated in the sense of GHSZ. The other GSZ states are not rotationally invariant, although they obey other group properties that reflect their degree of polarization. For example, the GHZ states given by equation (28) and their evolved states are not invariant under rotations, but they are invariant under the action of the group of order eight $Z_2 \times Z_2 \times Z_2$.

It might be asked if there are higher dimensional states that are also rotationally invariant. Indeed, the Fermi-Dirac state is (uniquely) invariant under the action of the $SL(n, C)$ group, which a fortiori means that it is invariant for all subgroups of $SL(n, C)$, including the rotation group. It should be noted that for two-dimensional subspaces of $C^n$, there are two rotationally invariant states and not one. This arises from the mathematical properties associated with Clifford algebras. In general, if $\boldsymbol{u}$ and $\boldsymbol{v}$ are two vectors in $C^n$ then we can define a Clifford product by

$$\mathbf{uv} = \mathbf{u}.\mathbf{v} + \mathbf{u} \wedge \mathbf{v}.$$

The first term corresponds to an inner product, which can be defined for any pair of vectors, and such a product is always rotationally invariant. However, there is no inner product for three vectors $\mathbf{u}, \mathbf{v}, \mathbf{w}$, meaning that $\mathbf{u}, \mathbf{v}, \mathbf{w}$, is not defined. In contrast, outer (wedge) products are defined over $n$-dimensions and correspond geometrically to volume (area in 2-dim), preserved under rotations. This leads to the invariance of the Fermi-Dirac state under $SL(n, C)$ as proven in the following theorem:

**Theorem 3.** *Let* $V = V_1 \otimes \cdots \otimes V_n$*, where each* $V_i$ *is an n-dimensional vector space, and* $T = T_1 \otimes \cdots \otimes T_n$ *where for each* $i, j, T_i = T_j$ *and* $T_i$ *is a linear operator on* $V_i$*. Let*

$$v \quad \equiv \quad v_1 \wedge v_2 \wedge \ldots \wedge v_n$$

$$= \begin{pmatrix} v_{11} \\ \vdots \\ v_{n1} \end{pmatrix} \wedge \begin{pmatrix} v_{12} \\ \vdots \\ v_{n2} \end{pmatrix} \wedge \ldots \wedge \begin{pmatrix} v_{1n} \\ \vdots \\ v_{nn} \end{pmatrix}$$

then for $v \neq 0$

$$Tv = v \Leftrightarrow T \in \otimes_1^n SL(n, \mathcal{C}).$$

*Fermi-Dirac statistics is invariant under the action of $SL(n, \mathcal{C})$.*

Proof: Let $\{\mathbf{e}_1, \mathbf{e}_2 \ldots \mathbf{e}_n\}$ be an orthonormal basis of $V_i$, then

$$v = \begin{pmatrix} v_{11} \\ \vdots \\ v_{n1} \end{pmatrix} \wedge \begin{pmatrix} v_{12} \\ \vdots \\ v_{n2} \end{pmatrix} \wedge \ldots \wedge \begin{pmatrix} v_{n1} \\ \vdots \\ v_{nn} \end{pmatrix}$$

$$= |v| \mathbf{e}_1 \wedge \mathbf{e}_2 \wedge \ldots \wedge \mathbf{e}_n,$$

where

$$|v| = \begin{vmatrix} v_{11} & v_{12} & \cdots & v_{1n} \\ v_{21} & v_{22} & \cdots & v_{2n} \\ \vdots & \vdots & \vdots & \vdots \\ v_{n1} & v_{n2} & \cdots & v_{nn} \end{vmatrix}.$$

The linearity of $T$ gives

$$Tv = |v| T_1 \mathbf{e}_1 \wedge T_2 \mathbf{e}_2 \wedge \ldots \wedge T_n \mathbf{e}_n$$

$$= |v| \begin{pmatrix} t_{11} \\ \vdots \\ t_{n1} \end{pmatrix} \wedge \begin{pmatrix} t_{12} \\ \vdots \\ t_{n2} \end{pmatrix} \wedge \ldots \wedge \begin{pmatrix} t_{n1} \\ \vdots \\ t_{nn} \end{pmatrix}$$

$$= |v| |T_1| \mathbf{e}_1 \wedge \mathbf{e}_2 \wedge \ldots \wedge \mathbf{e}_n, \qquad T_1 = T_2 = \cdots = T_n.$$

Therefore, since $v \neq 0$ implies $|v| \neq 0$ then

$$Tv = v \Rightarrow |T_1| = 1 \quad \text{and} \quad T_1 \in SL(n, \mathcal{C}).$$

Conversely

$$T_1 \in SL(n, \mathcal{C}) \Rightarrow Tv = v.$$

This proves the theorem. It is also shown in [4] that $v$ is unique.

## Acknowledgments

I would like to thank Andràs Kovàcs of Broadbit Energy Technologies (Finland) for having read the final manuscript and for offering valuable suggestions.

## Author Contributions

The author did all the research work for this study.

**Competing Interests**

The author has declared that no competing interests exist.

**References**

1. O'Hara P. Rotational invariance and the spin-statistics theorem. Found Phys. 2003; 33: 1349-1368.
2. O'Hara P. Entanglement—A Higher Order Symmetry. Phys Scie Forum. 2023; 7: 4.
3. O'Hara P. Entanglement, Microcausality and Gödel's Theorem. J Phys Conf Ser. 2023; 2482: 012013.
4. O'Hara P. A generalized spin statistics theorem. J Phys Conf Ser. 2017; 845: 012030.
5. O'Hara P. The Einstein-Podolsky-Rosen paradox and SU (2) relativity. J Phys Conf Ser. 2019; 1239: 012021
6. Bennett CH, DiVincenzo DP, Smolin JA, Wootters WK. Mixed-state entanglement and quantum error correction. Phys Rev A. 1996; 54: 3824.
7. Greenberger DM, Horne MA, Shimony A, Zeilinger A. Bell's theorem without inequalities. Am J Phys. 1990; 58: 1131-1143.
8. Norte RA, Forsch M, Wallucks A, Marinković I, Gröblacher S. Platform for measurements of the Casimir force between two superconductors. Phys Rev Lett. 2018; 121: 030405.
9. Huang H, Karan S, Padurariu C, Kubala B, Cuevas JC, Ankerhold J, et al. Universal scaling of tunable Yu-Shiba-Rusinov states across the quantum phase transition. Commun Phys. 2023; 6: 214.
10. Karan S, Huang H, Padurariu C, Kubala B, Theiler A, Black-Schaffer AM, et al. Superconducting quantum interference at the atomic scale. Nat Phys. 2022; 18: 893-898.
11. Aspect A, Dalibard J, Roger G. Experimental test of Bell's inequalities using time-varying analyzers. Phys Rev Lett. 1982; 49: 1804.
12. Gezerlis A, Carlson J. Strongly paired fermions: Cold atoms and neutron matter. Phys Rev C. 2008; 77: 032801.
13. Milloni P. The Quantum Vacuum. San Diego, CA: Academic Press; 1994. pp. 43-46, 98-105.
14. Chapovsky PL. Bose-Einstein condensation of rubidium atoms. JETP lett. 2012; 95: 132-136.
15. Georgescu I. 25 years of BEC. Nat Rev Phys. 2020; 2: 396.
16. Božović I, Levy J. Pre-formed Cooper pairs in copper oxides and $LaAlO_3$—$SrTiO_3$ heterostructures. Nat Phys. 2020; 16: 712-717.
17. Donley EA, Claussen NR, Thompson ST, Wieman CE. Atom–molecule coherence in a Bose–Einstein condensate. Nature. 2002; 417: 529-533.
18. Kovàcs A, Vassallo G. High-temperature superconductivity is the catalyzed Bose-Einstein condensation of electrons. András Kovács; 2023. pp. 17-19.
19. Bardeen J. Electron-phonon interactions and superconductivity. In: Cooperative Phenomena. Berlin and Heidelberg: Springer; 1973. p. 67.
20. Orlando TP, Delin KA. Foundations of applied superconductivity. Prentice Hall; 1991.
21. Ran S. Unboxing a New Spin-Triplet Superconductor [Internet]. Gaithersburg, MD: NIST; 2019. Available from: www.nist.gov/blogs/taking-measure/unboxing-new-spin-triplet-superconductor.

22. Piestrup MA, Adelphi Technology Inc. Enhanced Correlated-Charge Field Emission. Redwood City, CA: Adelphi Technology Inc.; 1998.

23. Ran S, Eckberg C, Ding QP, Furukawa Y, Metz T, Saha SR, et al. Nearly ferromagnetic spin-triplet superconductivity. Science. 2019; 365: 684-687.

24. Laing M. The three forms of molecular oxygen. J Chem Educ. 1989; 66: 453.

25. NIST. Strontium Unlocks the Quantum Secrets of Superconductivity [Internet]. Encinitas, CA: SciTech Daily; 2024. Available from: https://scitechdaily.com/strontium-unlocks-the-quantum-secrets-of-superconductivity/?utm_source=ground.news&utm_medium=referral

26. Young DJ, Chu A, Song EY, Barberena D, Wellnitz D, Niu Z, et al. Observing dynamical phases of BCS superconductors in a cavity QED simulator. Nature. 2024; 625: 679-684.